\documentclass[12pt]{article}
\usepackage{graphicx}
\usepackage{hyperref}
\usepackage{placeins}

\newcommand\pubnumber{SNSN-323-63}
\newcommand\pubdate{\today}

\def\institute{(a) Georg-August-Universit\"at G\"ottingen, G\"ottingen, Germany\\
(b) Universita degli Studi di Udine, Udine, Italy\\
(c) The University of Melbourne, Parkville, Australia \\
(d) Ludwig-Maximilians-Universit\"at M\"unchen, Germany \\
(e) University of Sussex, Brighton, United Kingdom\\
(f) Yale University, New Haven, CT USA\\
}

\usepackage{lineno}

\def\Title#1{\begin{center} {\Large #1 } \end{center}}
\def\Author#1{\begin{center}{ \sc #1} \end{center}}
\def\Address#1{\begin{center}{ \it #1} \end{center}}

\newcommand\pubblock{\rightline{\begin{tabular}{l} \pubnumber\\
         \pubdate  \end{tabular}}}
\newenvironment{Abstract}{\begin{quotation}  }{\end{quotation}}
\newenvironment{Presented}{\begin{quotation} \begin{center} 
             PRESENTED AT\end{center}\bigskip 
      \begin{center}\begin{large}}{\end{large}\end{center} \end{quotation}}

\def\beq{\begin{equation}}
\def\eeq#1{\label{#1}\end{equation}}
\def\eeqn{\end{equation}}

\def\beqa{\begin{eqnarray}}
\def\eeqa#1{\label{#1}\end{eqnarray}}
\def\eeqan{\end{eqnarray}}

\let\bar=\overbar

\def\Dslash{\not{\hbox{\kern-4pt $D$}}}
\def\dslash{\not{\hbox{\kern-2pt $\del$}}}

\def\msb{{\bar{\ssstyle M \kern -1pt S}}}

\begin{document}
\begin{titlepage}
\pubblock

\vfill
\Title{Top SciComm: Communicating ATLAS Top Physics Results to the Public}
\vfill
\Author{Clara Nellist$^{a}$, Katarina Anthony$^{b}$, Steven Goldfarb$^{c}$, Sascha Mehlhase$^{d}$, Kate Shaw$^{e}$, Savannah Jennifer Thais$^{f}$, Emma Ward$^{b}$}
\Address{\institute}
\vfill
\begin{Abstract}
An essential component of the long-term success of scientific research is communicating the methodology and significance of new results to the wider public. Utilising various social media platforms is a vital tool for this endeavour. Over the years, there have been a number of important results on top physics released by the ATLAS Collaboration. These have been communicated through audience-tailored content, including ATLAS physics briefings, videos, and press statements, amongst others. The ATLAS Collaboration has continued to adapt its communication strategy to match the social media evolution, producing content specifically targeting this emerging audience, the effect of which has been explored.
\end{Abstract}
\vfill
\begin{Presented}
$11^\mathrm{th}$ International Workshop on Top Quark Physics\\
Bad Neuenahr, Germany, September 16--21, 2018
\end{Presented}
\vfill
\small
\textcopyright 2019 CERN for the benefit of the ATLAS Collaboration.\\
Reproduction of this article or parts of it is allowed as specified in the CC-BY-4.0 license.
\end{titlepage}
\def\thefootnote{\fnsymbol{footnote}}
\setcounter{footnote}{0}

\section{Introduction}

An essential component of the long-term success of scientific research is communicating the methodology and significance of new results to the wider public. Utilising various social media platforms is a vital tool for this endeavour.
The main aims of the ATLAS social media strategy are to:
\begin{itemize}
\item Show not just what science is being done, but also how it is done. This includes the scientists behind it, and the international nature of the Collaboration.
\item Make ATLAS research more accessible to a wider audience.
\item Share information on ATLAS research with other physicists.
\item Reach future ATLAS students / researchers.
\item Direct people to the ATLAS website~\footnote{\url{www.atlas.cern}}, where content is available in more depth, if it is not possible to provide all the content in the platform. Furthermore, this also introduces them to other related content.
\end{itemize}

\section{Tailored content}

\begin{figure}[htb]
\centering
\includegraphics[width=0.44\textwidth]{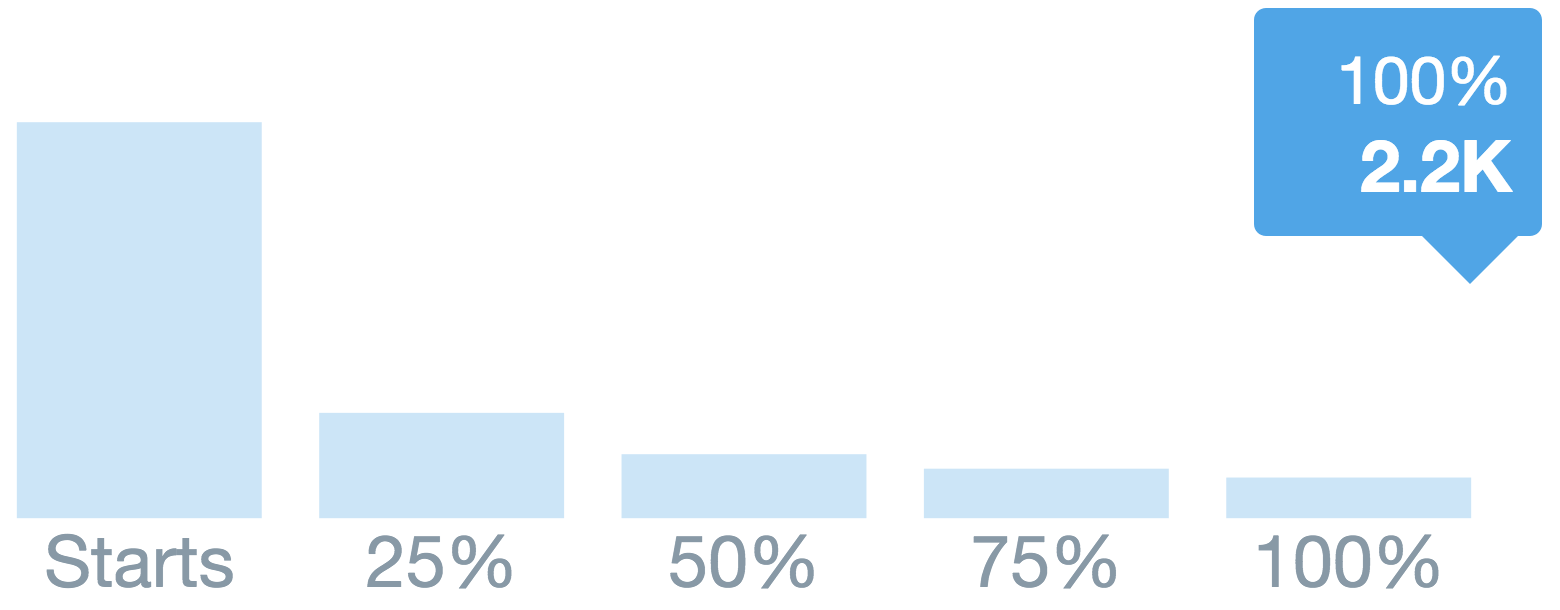}
\caption{Video retention on the @ATLASExperiment Twitter account for August 2018. The first bin labelled `starts' indicates the number of people who watched the video for longer than 2 seconds. Each subsequent bin is the number of people who reached that percentage of the video. Numbers correct as of September 2018.}
\label{fig:twitter-retention}
\end{figure}

In recent years, an increasing effort has been devoted to the goal of tailoring the content to the specific social media platform being used. Various methods are implemented to keep users on the platform and grab their attention. Across our various platforms we need to cater to the algorithms that select which content is given priority to be shown to users to ensure our content is more widely distributed. Short videos of approximately one minute have been created with concise content since audience retention longer than this time drops off significantly. The aim is also to have the main message delivered within three seconds, while retaining accuracy, to endeavour to get our message across even with low audience retention, as shown in Figure~\ref{fig:twitter-retention}.

The use of captions for accessibility is important to ensure that we are not excluding various demographics of the audience. This is also useful since many users are no longer choosing to have the sound on during videos if, for example, they are in a public place. The implementation of a 1:1 video ratio for Facebook and Instagram is shown to increase engagement with audiences~\footnote{\url{https://blog.bufferapp.com/square-video-vs-landscape-video} Accessed 22nd January 2019.}. The hypothesis is that this video ratio fills up more of the screen on a mobile device held in a portrait orientation, and is therefore more likely to lead to engagement with the user.\\

\begin{figure}[htb]
\centering
\begin{minipage}[c]{0.3\textwidth}
\includegraphics[width=\textwidth]{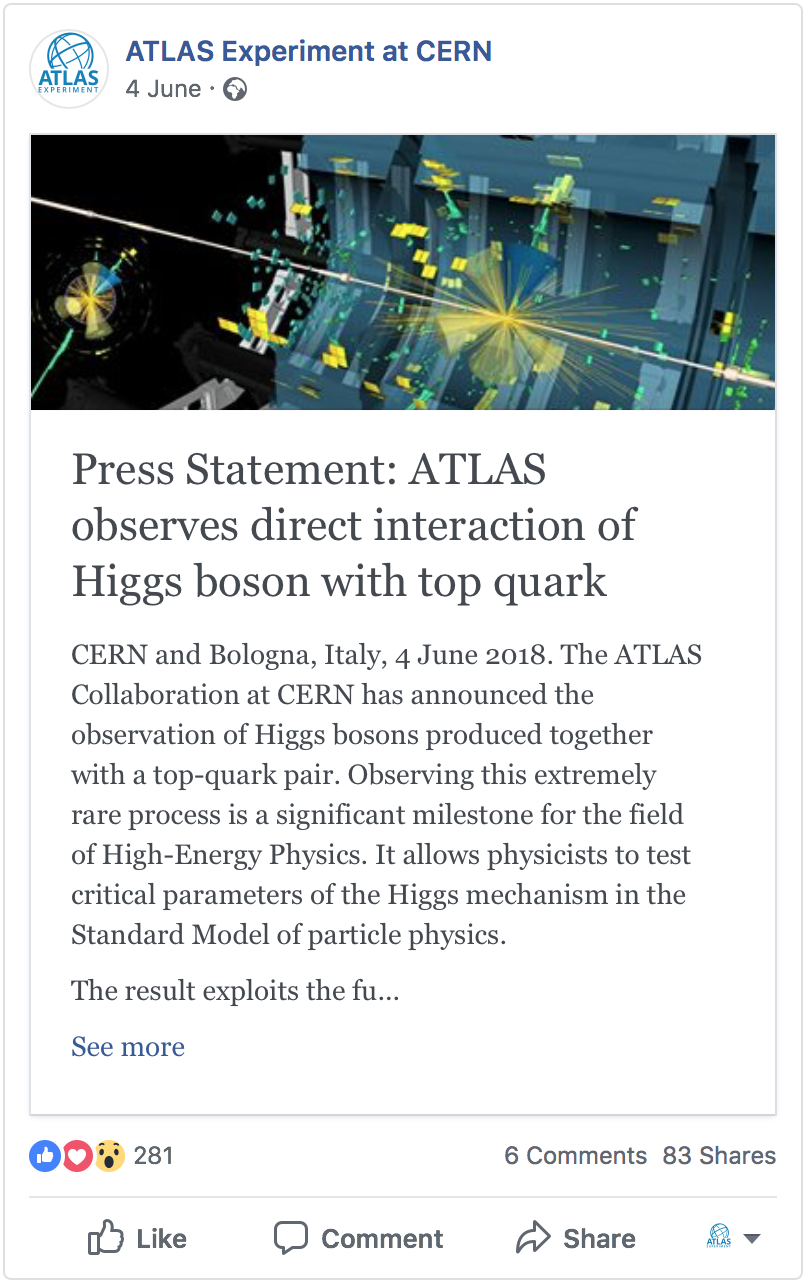}
\end{minipage}
\begin{minipage}[c]{0.5\textwidth}
\caption{An ATLAS press statement that was published on the ATLAS Facebook page directly as a Note. As of September 2018, 19,543 people had been reached, including users reached through shares by other accounts.}
\label{fig:fb-note}
\end{minipage}
\end{figure}

Longer text documents that were previously only published on the ATLAS website, and then linked to on our social media platforms, are now also posted as Facebook Notes. An example Note is shown in Figure~\ref{fig:fb-note}, for the press statement for the observation of a Higgs boson produced in association with a top-quark pair~\cite{ATLAS:ttH}.

Instagram Stories, short images or videos that are usually only visible on the platform for 24 hours, have been implemented to highlight posts. After a new ATLAS top-physics result has been published, there are various ways it can be publicised on our platforms. Further details specific to a platform will be illustrated in the follow subsections.

\subsection{Twitter}

\begin{figure}[htb]
\centering
\includegraphics[height=0.26\textheight]{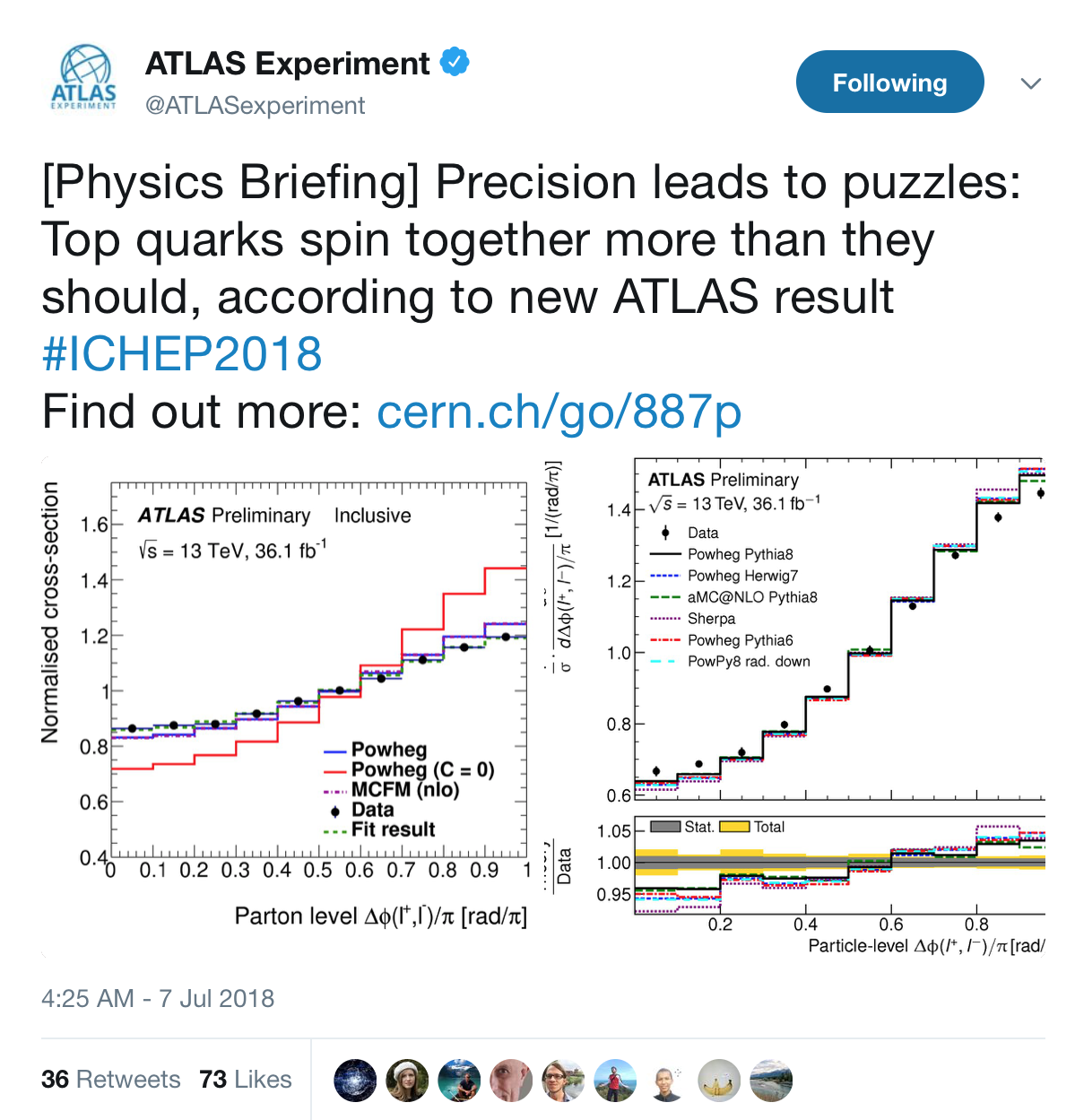}
\includegraphics[height=0.26\textheight]{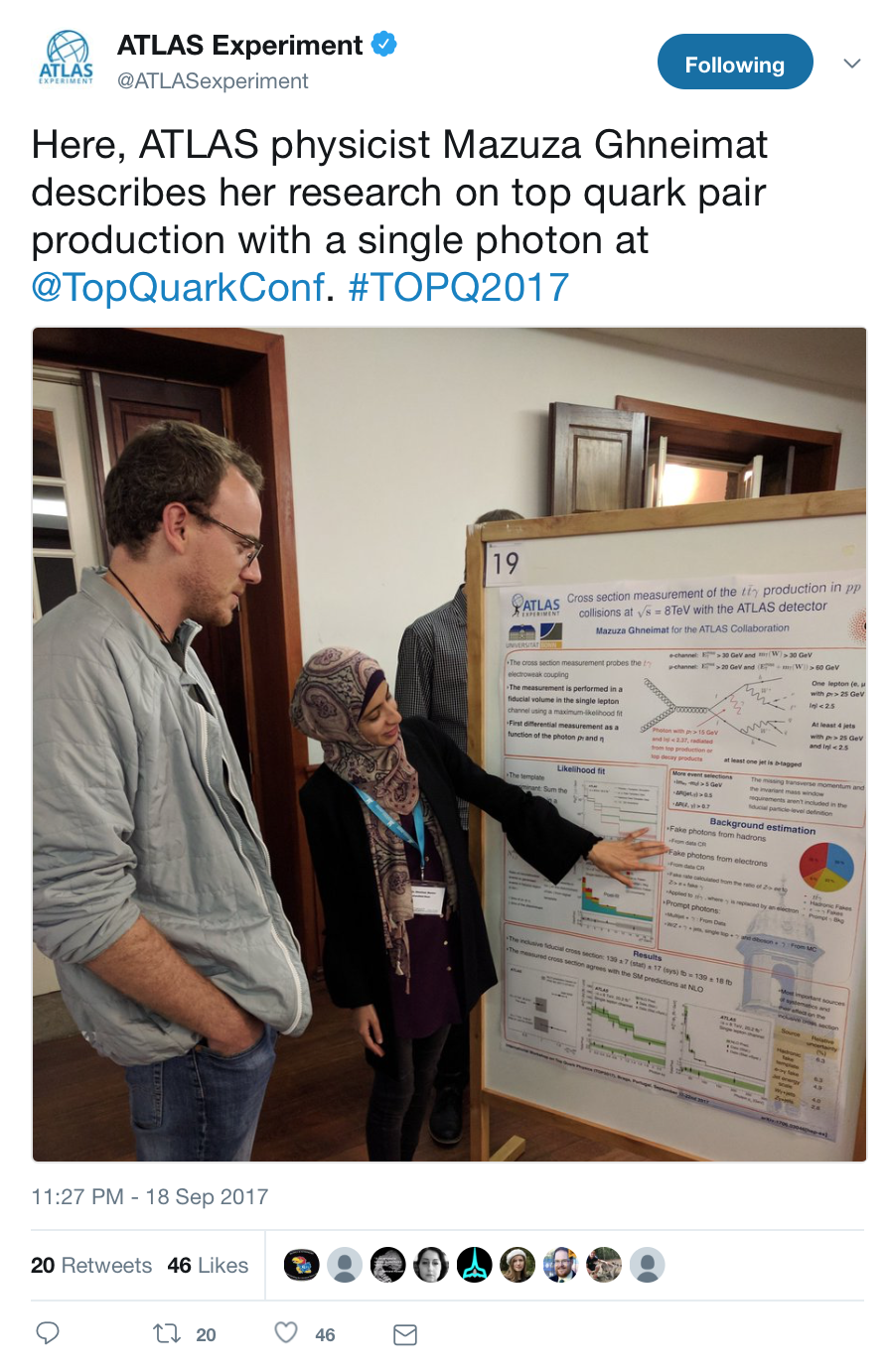}
\caption{Example tweets posted on the ATLAS Twitter account about a physics briefing (left), and an announcement about the TOP2017 workshop (right). Numbers correct as of September 2018.}
\label{fig:tweets}
\end{figure}

In Figure~\ref{fig:tweets}, there are two example tweets. The first, on the left, is advertising a new physics briefing that has been posted on the ATLAS website. Two plots from the paper which have been used in the briefing are included in the tweet as it has been shown that there is increased engagement on Twitter when images or videos are included. The second example, on the right, is an announcement about the start of the TOP2017 Workshop, indicating the Twitter account and hashtag that users can follow to read about the workshop.

\subsection{Facebook}

\begin{figure}[htb]
\centering
\begin{minipage}[c]{0.57\textwidth}
\includegraphics[width=1\textwidth]{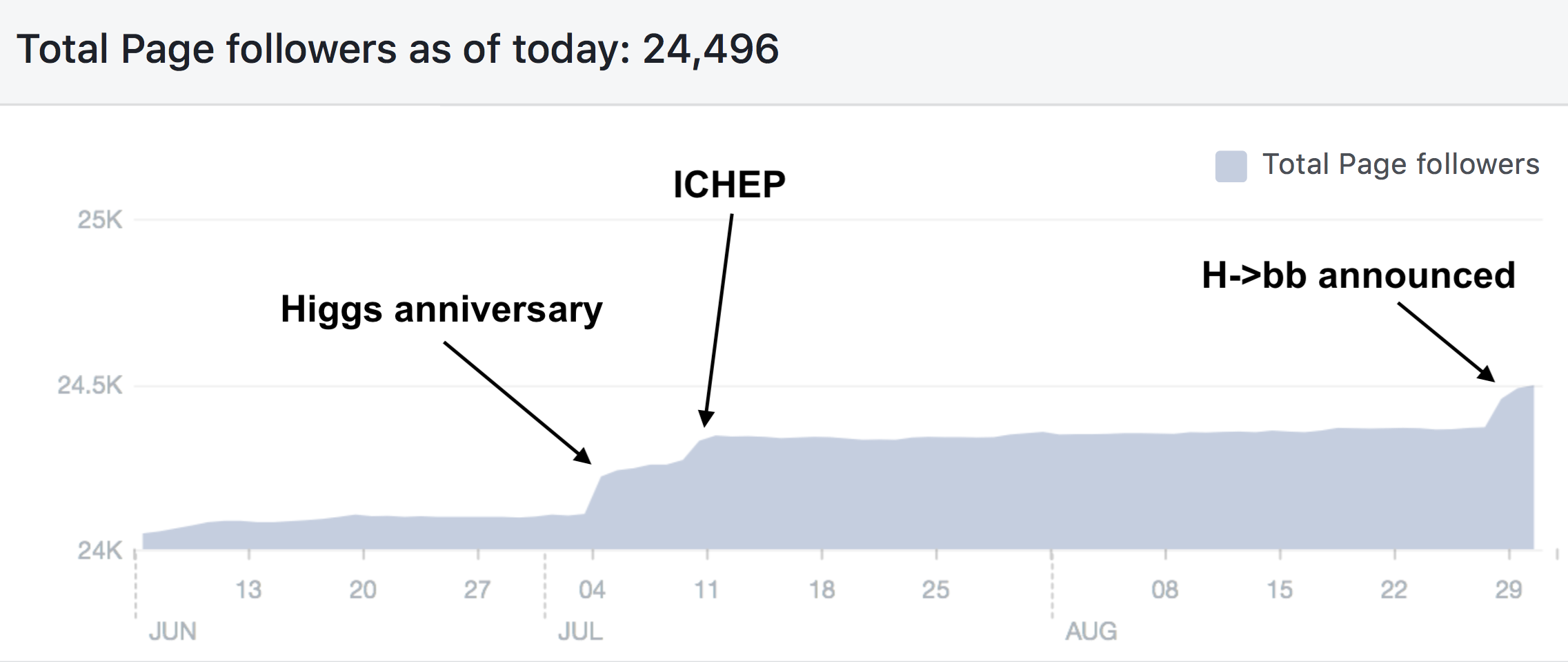}
\end{minipage}
\begin{minipage}[c]{0.42\textwidth}
\caption{Total ATLAS Experiment Facebook Page followers as of September 2018. The sudden increases in followers are labelled in the figure as to the most likely event to increase the exposure of new people to the account.}
\label{fig:fb1}
\end{minipage}
\end{figure}

The ATLAS Facebook account has approximately 25 thousand followers, as of September 2018. This number increases over time as content is posted on the site. However, certain events or posts result in a larger increase of followers, suggesting that this content is being shared more widely by our existing audience and the engagement is positive enough that a certain fraction chooses to follow the account. This is illustrated in Figure~\ref{fig:fb1}.

\begin{figure}[htb]
\centering
\begin{minipage}[c]{0.66\textwidth}
\includegraphics[width=1\textwidth]{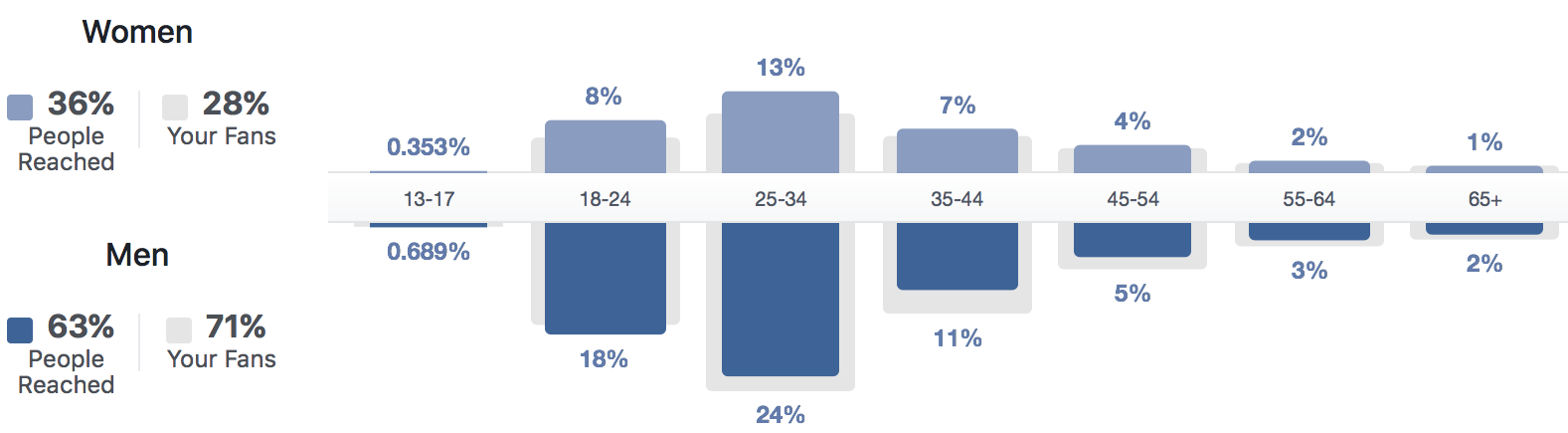}
\end{minipage}
\begin{minipage}[c]{0.33\textwidth}
\caption{Distribution of followers for the ATLAS Facebook Page by age and gender. Numbers correct as of September 2018.}
\label{fig:fb2}
\end{minipage}
\end{figure}

The breakdown of followers on the ATLAS Facebook Page by age and gender can be seen in Figure~\ref{fig:fb2}. Facebook currently only reports binary gender of followers. This shows that there is scope to further engage women on the Page.

\subsection{Instagram}

\begin{figure}[htb]
\centering
\begin{minipage}[c]{0.21\textwidth}
\includegraphics[width=1\textwidth]{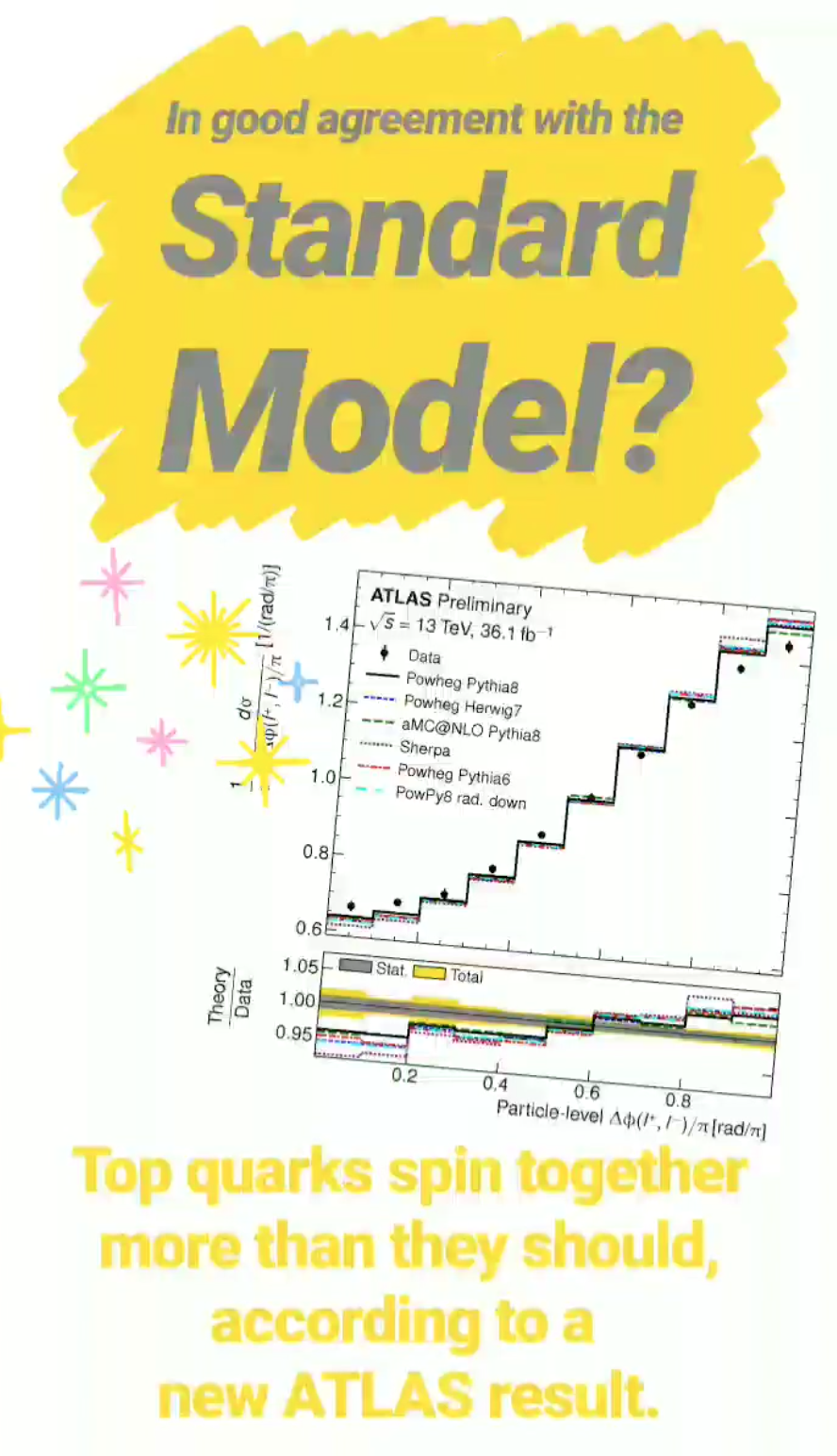}
\end{minipage}
\begin{minipage}[c]{0.4\textwidth}
\caption{Example Instagram Story highlighting the top-quark pair spin-correlations result as part of the ICHEP promotion.}
\label{fig:Insta}
\end{minipage}
\end{figure}

Stories on Instagram is a feature that allows the user to post images or a video of less than 15 seconds, which can only be viewed for 24 hours\footnote[1]{There is also the option to feature stories on the account.}. These can be more whimsical than standard posts, and the time limit encourages users to actively visit the platform frequently. An example story highlighting the top-quark pair spin correlations result~\cite{ATLAS:2018rgl} as part of the ICHEP promotion on the ATLAS Experiment account can be seen in Figure~\ref{fig:Insta}.

\FloatBarrier

\section{Conclusions}

Communication of scientific results via various social media platforms is a dynamic method to reach a growing audience. The ATLAS Collaboration has continued to adapt its communication strategy to match the social media evolution, producing content specifically targeting this emerging audience, the effect of which has been explored in these proceedings. The changes have been found to be positive and new methods to communicate effectively will continue to be explored.





\end{document}